\documentclass{article}

\usepackage{arxiv}

\usepackage[utf8]{inputenc} % allow utf-8 input
\usepackage[T1]{fontenc}    % use 8-bit T1 fonts
\usepackage{hyperref}       % hyperlinks
\usepackage{url}            % simple URL typesetting
\usepackage{booktabs}       % professional-quality tables
\usepackage{amsfonts}       % blackboard math symbols
\usepackage{nicefrac}       % compact symbols for 1/2, etc.
\usepackage{microtype}      % microtypography
\usepackage{lipsum}
\usepackage{graphicx}
\graphicspath{ {./images/} }
\usepackage{amsmath}
\usepackage{amsmath,amssymb,amsfonts}
\usepackage{bm}
\usepackage[linesnumbered,ruled]{algorithm2e}
\usepackage{color}
\usepackage{multirow}
\usepackage{makecell}
\usepackage{booktabs}
\usepackage[T1]{fontenc}
\usepackage{lineno}
\usepackage{soul}
\usepackage{pifont}
\usepackage{multirow}

\title{A Multi-physics Alternating Coupled Inversion Using Gravity and Full Waveform Data in Salt Dome}

\author{
	Siyuan Dong \\
	School of Information and Communications Engineering\\
	Xi'an Jiaotong University\\
	Xi'an, Shanxi, China \\
	\texttt{sydong24@stu.xjtu.edu.cn} \\
	\And
	Jinghuai Gao \\
	School of Information and Communications Engineering\\
	Xi'an Jiaotong University\\
	Xi'an, Shanxi, China \\
	\texttt{jhgao@mail.xjtu.edu.cn} \\
	\And
	Yunduo Li \\
	School of Information and Communications Engineering\\
	Xi'an Jiaotong University\\
	Xi'an, Shanxi, China \\
	\texttt{yunduo\_li@stu.xjtu.edu.cn} \\
	\And
	Zhaoqi Gao \\
	School of Information and Communications Engineering\\
	Xi'an Jiaotong University\\
	Xi'an, Shanxi, China \\
	\texttt{zq\_gao@xjtu.edu.cn} \\
	\And
	Baohai Wu \\
	School of Information and Communications Engineering\\
	Xi'an Jiaotong University\\
	Xi'an, Shanxi, China \\
	\texttt{baohaiwu@163.com} \\
	\And
	Feng Liu \\
	School of Electronic Information and Electrical Engineering\\
	Shanghai Jiao Tong University\\
	Shanghai, China \\
	\texttt{liufeng2317@sjtu.edu.cn} \\
}

\begin{document}
\maketitle
\begin{abstract}
\label{sec:abs}
Complex salt geometries and strong velocity contrasts pose significant challenges for velocity model building and subsalt imaging. Although full waveform inversion (FWI) provides high‐resolution velocity models, its performance strongly depends on the accuracy of initial model. On the other hand, gravity focusing inversion (GFI) can recover compact density distributions and provide reliable long‐wavelength structural information for seismic exploration, but it suffers from poor depth resolution and inherent non‐uniqueness. To better invert salt structure by leveraging the complementary advantages of full waveform and gravity data, we propose a multi-physics alternating coupled inversion strategy for salt dome model. The proposed strategy mainly includes three parts. First, we perform FWI using a simple layered velocity model to obtain preliminary velocity updates and extract the salt top boundary. Second, this structural information is used as a constraint in GFI to recover a compact salt density distribution beneath the salt top. Third, the resulting salt geometry is used to construct an improved velocity model for the next stage of FWI. Through iterative alternation, FWI provides reliable structural constraints for GFI, while GFI supplies a more reasonable macroscopic salt model for FWI, effectively mitigating the strong dependence on the initial model. In addition, a depth-varying density contrast is introduced in GFI to better represent sediment compaction effects. Compared with unconstrained GFI and conventional FWI using a horizontally layered initial model, the proposed method effectively improves both velocity and density reconstruction in the modified BP salt model and SEG/EAGE salt model.
\end{abstract}

% keywords can be removed

\section{Introduction}
\label{sec:intro}
Salt structures, characterized by low permeability and strong plastic flow, can serve both as high-quality seals and lateral barriers, while also inducing salt-related fault-fold structures and flank traps. These contribute to the formation of diverse types of effective traps and control hydrocarbon accumulation \cite{hudec2007terra}. Consequently, the subsalt and near-flank areas often become favorable exploration zones for giant oil and gas fields \cite{tang2004salt, qiu2020recognition}. Therefore, constructing a high-precision velocity model in salt dome is a critical prerequisite for reliable subsalt imaging and reservoir evaluation \cite{leveille2011subsalt, wang2019full, luo2023salt}.

As a high-resolution velocity modeling reconstruction methods, FWI \cite{tarantola1984inversion, tarantola1986strategy} leverages the diverse information contained within seismic waveforms and exhibits a strong capability for reconstructing subsurface parameters. It has shown significant application potential for velocity modeling in complex structural areas \cite{virieux2010overview, gao2017novel, gao2022self, du2025structurally}. However, applying FWI in salt dome still faces numerous challenges. The complex geometry of salt bodies and the strong velocity contrast with surrounding sediments significantly affect seismic wave propagation. This leads to imaging distortion and increases uncertainty in subsalt structural interpretation. Moreover, such features introduce extreme nonlinearity into the seismic wavefield. This nonlinearity makes FWI highly dependent on an accurate initial velocity model. When the initial model is inaccurate, FWI is prone to cycle-skipping issues, causing the inversion to converge to erroneous local minima and compromising the reliability of the results. Therefore, relying solely on full waveform data for velocity inversion in complex salt dome environments still presents certain limitations. In recent years, the use of multiple geophysical datasets to invert for a consistent subsurface model has gained increasing attention in geophysics \cite{colombo2018coupling, lin2018joint, zhdanov2023advanced, hu2023deep}. As a developing trend in exploration methods, incorporating other geophysical information to constrain FWI is an effective way to enhance its stability and reliability \cite{treister2017full, feng2017joint, jiang2020detecting, guo2025deep}.

GFI is widely applied in salt dome characterized by significant density contrasts, due to its ability to produce compact and focused density results \cite{gao2017research, lin2019gramian, chen20233, xian2024recovering}. Under the assumption of uniform sedimentary layer density, the density of salt body is typically lower than that of the surrounding sediments, exhibiting a clear negative anomaly. Furthermore, GFI has a relatively low dependence on the initial model and can provide long-wavelength structural information for seismic exploration, offering unique advantages in building macroscopic models. However, gravity data acquired from passive source surveys inherently lack depth resolution, which leads to limited depth information in the inversion and introduces significant non-uniqueness \cite{dong2026enhanced}. Therefore, combining full waveform data, which carry rich depth information, with gravity data through joint or cooperative inversion holds the potential to leverage the complementary strengths of both data types. 

In recent years, several studies have attempted to integrate gravity data with FWI to improve velocity and density inversion in salt dome. \cite{pontius2016joint} jointly implemented two-dimensional (2-D) acoustic FWI and gravity inversion in a synthetic salt model. \cite{jiang20203} successfully performed three-dimensional (3-D) joint inversion using full waveform and airborne gravity gradient data in both synthetic and field salt models. In both studies, the initial velocity model for FWI is obtained from a smoothed version of the true model or from traveltime tomography, without leveraging the characteristics of gravity field to build a macroscopic model for FWI. Moreover, the complex objective function in joint inversion significantly increases the computational and parameter-tuning costs. \cite{silva2020cooperative, silva2022cooperative} proposed a cooperative strategy between full waveform and gravity to better control the inversion of each data type individually. In this strategy, a smoothed true velocity model is first used as the initial velocity model. The velocity model updated by FWI is then converted into a density model for gravity inversion via Gardner's density–velocity relationship. Subsequently, Gardner's relation is used again to convert the updated density model back into a velocity model for the next round of FWI. However, using a smoothed true velocity model as the initial model is overly idealized and deviates significantly from complex real geological conditions.

Additionally, considering the effects of compaction in reality, the density of sedimentary layers generally increases with depth. This results in a positive density contrast for the salt body in the shallow section, leading to a positive gravity anomaly in the surface data. In the deeper section, the density contrast becomes negative, corresponding to a negative gravity anomaly at the surface. Between these zones, there exists a region where the density of the salt body equals that of the surrounding sediments, known as the null zone. Within the null zone, the salt body does not generate any surface gravity anomaly. Furthermore, the positive and negative anomalies produced by the shallow and deep sections of the salt body may partially cancel each other in the surface gravity data, a phenomenon referred to as the annihilation \cite{gibson1998geologic}. Under such circumstances, conventional GFI struggles to address this issue. To tackle this, \cite{krahenbuhl2006inversion} introduced the true salt top structure as a constraint and applied a binary inversion method, achieving reasonable recovery of the salt geometry near the null zone. However, enforcing a discrete 0–1 density contrast renders traditional derivative-based minimization techniques entirely inapplicable, leading to a substantial increase in algorithmic complexity and computational cost. \cite{li2016level} adopted a level-set method to address the imaging problem in the null zone of salt bodies, representing the salt boundary as the zero level set of a level-set function, thereby enabling optimization via gradient-based methods. \cite{deng2025level} further extended this approach to joint FWI and gravity inversion, achieving structural consistency between velocity and density parameters through a shared interface. Nevertheless, the level-set method still suffers from high computational cost and complex weight tuning.

To better exploit the complementary nature of gravity and full waveform data and facilitate their mutual constraint, we propose an alternating coupled inversion method for salt dome. Through an alternating iterative procedure, the method gradually improves the reliability of the inverted velocity and density models. Specifically, a simple horizontally layered model is first used as the initial velocity model for FWI. The top boundary of the salt body is then extracted from the FWI result and introduced as a structural constraint in the GFI, where density updates are confined to the region below the salt top. This yields a compact density model that closely conforms to the salt top boundary. Based on the salt geometry obtained from GFI, a more realistic initial velocity model for the salt body is constructed and embedded into the background velocity field for the next stage of FWI. In this alternating coupled inversion framework, FWI provides reliable structural constraints for GFI, while GFI supplies a more accurate initial salt model for FWI. This effectively mitigates inversion instability caused by inaccurate initial models. Compared with unconstrained GFI and conventional FWI, the proposed strategy progressively improves the accuracy of both velocity and density inversion in salt dome.

This paper is organized as follows. In the "Theory" section, we first review the theory of FWI and introduce an algorithm for salt top boundary extraction. Next, we present the conventional GFI assuming a constant density contrast, and then extend it to account for depth-varying density contrast. Finally, we detail the workflow of proposed alternating coupled inversion method. In "Numerical Examples" section, we apply the proposed method to modified BP salt model and SEG/EAGE salt model. Lastly, we summarize our study in the "Conclusion" section.

\section{Theory}
\label{sec:theroy}
\subsection{Full Waveform Inversion}
\label{subsec:FWI}
Constant density acoustic FWI is a highly nonlinear inverse problem that updates the P-wave velocity model ${\bf m}_s$ by minimizing the least-squares objective function $\mathcal{J} ({\bf m}_s)$:

\begin{linenomath*}
	\begin{align}\label{eq:fwi_base}
%		\varPhi ({\bf m}_s) = \frac{1}{2} \sum_{x_s} \sum_{x_r} \int_t \left[d_s^{\rm obs}(x_s,x_r,t) - d_s^{\rm syn}({\bf m}_s;x_s,x_r,t)\right]^2 dt,
		\mathcal{J} ({\bf m}_s) = \sum_{x_s} \Vert {\bf d}_s(x_s) - {\bf S} ({\bf m}_{s}, x_s) \Vert^2,
	\end{align}
\end{linenomath*}

%where $d_s^{\rm obs}(x_s,x_r,t)$ and $d_s^{\rm syn}({\bf m}_s;x_s,x_r,t)$ respectively represent the observed wavefield and simulated wavefield with respect to source location $x_s$, receiver location $x_r$, and time $t$
where ${\bf d}_s$ is observed wavefield, $\bf S$ is forward modeling operator, and $x_s$ is source location. The optimization in FWI can be generally expressed as: 

\begin{linenomath*}
	\begin{align}\label{eq:fwi_optim}
		{\bf m}_s^* = \mathop{\arg\min}\limits_{{\bf m}_s}[\mathcal{J} ({\bf m}_s) + \alpha \mathcal{R}({\bf m}_s)],
	\end{align}
\end{linenomath*}

where ${\bf m}_s^*$ represents the final estimated velocity, $\mathcal{R}({\bf m}_s)$ denotes the regularization term, and $\alpha$ is the corresponding damping coefficient. In this study, we use the second-order total variation (TV2) regularization, which preserves sharp boundaries while suppressing staircase effects \cite{dogan2011second}. Its expression is given as follows:

\begin{linenomath*}
	\begin{align}\label{eq:fwi_TV2}
		\mathcal{R}_{\text{TV2}}({\bf m}_s) = \alpha_x |{\bf D}_{xx}({\bf m}_s)| + \alpha_z |{\bf D}_{zz}({\bf m}_s)|,
	\end{align}
\end{linenomath*}

where ${\bf D}_{xx}$ and ${\bf D}_{zz}$ are the second-order difference operators in the horizontal and vertical directions, $\alpha_x$ and $\alpha_z$ are the corresponding weights. Here, the gradient of $\mathcal{J} ({\bf m}_s)$ with respect to ${\bf m}_s$ is obtained through automatic differentiation \cite{liu2025automatic}.

\subsection{Salt Top Boundary Extraction}
\label{subsec:Top_boundary}
The top of salt body is typically characterized by a significant vertical velocity contrast. Therefore, the core idea of boundary extraction is to identify the point corresponding to the maximum vertical velocity change for each trace within the window. First, we define the horizontal window $X = \left\{ {i:{x_{{\rm{min}}}} \le {x_i} \le {x_{{\rm{max}}}}} \right\}$ and vertical window $Z = \left\{ {k:{z_{{\rm{min}}}} \le {z_k} \le {z_{{\rm{max}}}}} \right\}$, where $x_i$ and $z_k$ denote the lateral and depth coordinates of the grid point $(i,k)$, respectively. $[x_{{\rm{min}}}, x_{{\rm{max}}}]$ and $[z_{{\rm{min}}}, z_{{\rm{max}}}]$ are the horizontal and vertical ranges used for salt top detection. Within this window, the magnitude of vertical velocity variation $\bf G{(x,z)}$ of ${\bf m}_s^*$ is computed as follows:

\begin{linenomath*}
	\begin{align}\label{eq:salt_top_cal}
		G{(x,z)_{i,k}} = \sqrt {(\frac{{\partial m_s^*}}{{\partial x}})_{i,k}^2 + (\frac{{\partial m_s^*}}{{\partial z}})_{i,k}^2}.
	\end{align}
\end{linenomath*}

Since the salt top is generally associated with a positive downward velocity increase, only points satisfying $({\partial m_s^*}/{\partial z})_{i,k} > 0$ are retained as valid candidates. To ensure lateral continuity of the extracted salt top boundary, a maximum inter-trace jump constraint $\sigma$ is introduced. If the picked depth index on the $(i-1)$-th trace is $k_{i-1}^*$, then the salt-top index $k_i^*$ on the $i$-th trace is determined by:

\begin{linenomath*}
	\begin{align}\label{eq:salt_k}
		k_i^* = \mathop{\arg\max}\limits_{\substack{k\in Z\\ |k-k_{i-1}^*|\le \sigma}} G{(x,z)_{i,k}}.
	\end{align}
\end{linenomath*}

Finally, the discrete picked boundary is smoothed to obtain a geologically plausible continuous salt top interface.

\subsection{Gravity Focusing Inversion}
\label{subsec:GFI}
In GFI, the objective function $\mathcal{J} ({\bf m}_g)$ based on the minimum support stabilizer \cite{last1983compact} with respect to the observed gravity field data ${\bf d}_{g}$ and the density model ${\bf m}_{g}$ is given as follows:

\begin{linenomath*}
	\begin{align}\label{eq:grainv_MS}
		\mathcal{J} ({\bf m}_g) = \Vert {\bf d}_g - {\bf A} {\bf m}_g \Vert^2 + \lambda \Vert {\bf W}_e {\bf m}_g \Vert^2,
	\end{align}
\end{linenomath*}

where $\bf A$ is the sensitivity matrix and ${\bf W}_e$ is a diagonal matrix with elements of $w_e(m_g)=(m_g^2+e^2)^{-1/2}$ along its main diagonal. $m_g$ is an element of ${\bf m}_g$, $e$ denotes the focusing factor, and $\lambda $ is the corresponding damping coefficient. In classical GFI, the density contrast is assumed to be constant, lower and upper density bounds $(\rho_{\rm {min}}, \rho_{\rm {max}})$ are set such that ${\bf m}_{g} \in [\rho_{\rm {min}}, \rho_{\rm {max}}]^N$. In this case, $\rho_{\rm {min}}$ represents the density contrast between the salt body and the sedimentary layer, and $\rho_{\rm {max}} = 0$. Additionally, we introduce model weighting matrix ${\bf W}_m = {\rm diag}({\bf A}^\top {\bf A})^{1/2}$ and data weighting matrix ${\bf W}_d = {\rm diag}({\bf A} {\bf A}^\top)^{1/2}$ to balance the contribution of model at different depths and to mitigate the skin effect \cite{portniaguine20023}. The inverse problem is then solved in the following weighted density parameter domain\cite{gao2017research}:

\begin{linenomath*}
	\begin{align}\left\{\begin{aligned}\label{eq:grainv_weightedspace}
			{\bf m}^w_g & = {\bf W}_e {\bf W}_m {\bf m}_g \\
			{\bf A}^w & = {\bf W}_d {\bf A} {\bf W}^{-1}_m {\bf W}^{-1}_e. \\
			{\bf d}^w_g & = {\bf W}_d {\bf d}_{g}
		\end{aligned}\right.\end{align}
\end{linenomath*}

Substituting Equation \ref{eq:grainv_weightedspace} into Equation \ref{eq:grainv_MS} yields the GFI objective function $\mathcal{J} ({\bf m}^w_g)$ for iterative weighted density model ${\bf m}^w_g$ updates:

\begin{linenomath*}
	\begin{align}\label{eq:grainv_FI}
		\mathcal{J} ({\bf m}^w_g) = \Vert {\bf d}^w_g - {\bf A}^w {\bf m}^w_g \Vert^2 + \lambda \Vert {\bf m}^w_g \Vert^2.
	\end{align}
\end{linenomath*}

In GFI, science the density contrast $\Delta\rho(z)$ varies with depth $z$, directly inverting for density ${\bf m}_{g}$ using conventional GFI (such as Equation \ref{eq:grainv_FI}) poses a significant challenge. To address this limitation, we introduce a salt indicator function ${\bm {\mathcal{X}}_g} \in [0, 1]^N$ and reparameterize the conventional linear gravity inverse problem ${\bf d}_g = {\bf A} {\bf m}_g$ as follows:

\begin{linenomath*}
	\begin{align}\label{eq:FI_DV}
		{\bf d}_g = {\bf \widetilde{A}} \bm {\mathcal{X}}_g,
	\end{align}
\end{linenomath*}

where ${\bf \widetilde{A}} = {\bf A} {\rm diag}(\Delta\rho(z))$. In this case, the inversion target is no longer the density ${\bf m}_{g}$ of the salt body but the dimensionless salt indicator function $\bm {\mathcal{X}}_g$. Additionally, $\bm {\mathcal{X}}_g$ values closer to 0 indicate a tendency toward the background sediments and values closer to 1 indicate a tendency toward the salt body. In this stage, the objective function $\mathcal{J} (\bm {\mathcal{X}}_g)$ is given as follows:

\begin{linenomath*}
	\begin{align}\label{eq:FI_loss_DV}
		\mathcal{J} (\bm {\mathcal{X}}_g) = \Vert {\bf d}_g - {\bf \widetilde{A}} \bm {\mathcal{X}}_g \Vert^2 + \lambda \Vert \widetilde{\bf W}_e \bm {\mathcal{X}}_g \Vert^2,
	\end{align}
\end{linenomath*}

where $\widetilde{\bf W}_e$ is a diagonal matrix with elements of $\widetilde{w}_e(\mathcal{X}_g)=({\mathcal{X}_g}^2+e^2)^{-1/2}$ along its main diagonal, and $\mathcal{X}_g$ is an element of $\bm {\mathcal{X}}_g$. Since $\Delta\rho(z)$ approaches zero in the mid-depth region of the salt body, further constructing model and data weighting matrices based on ${\bf \widetilde{A}}$ would degenerate the corresponding weights in this region. This leads to numerical instability and impairing the imaging capability in the central part of the salt body. Therefore, in GFI that accounts for depth-varying density contrast, we adopt an inversion formulation without explicit sensitivity weighting. Instead, we rely primarily on the top boundary constraints provided by FWI and focusing regularization to ensure the continuity of the salt body. 

In this study, the gradients of $\mathcal{J} ({\bf m}^w_g)$ and $\mathcal{J} (\bm {\mathcal{X}}_g)$ with respect to ${\bf m}^w_g$ and $\bm {\mathcal{X}}_g$ are computed using the conjugate gradient (CG) algorithm \cite{portniaguine1999focusing}:

\begin{linenomath*}
	\begin{align}\label{eq:gradient_FI}
		\frac{\partial \mathcal{J} ({\bf m}^w_g)}{\partial {\bf m}^w_g} =  ({\bf A}^{w\top} {\bf A}^w + \lambda {\bf I}){\bf m}^w_g -{\bf A}^{w\top} {\bf d}^w_g,
	\end{align}
\end{linenomath*}

\begin{linenomath*}
	\begin{align}\label{eq:gradient_FI_DV}
		\frac{\partial \mathcal{J} (\bm {\mathcal{X}}_g)}{\partial \bm {\mathcal{X}}_g} =  ({\bf \widetilde{A}}^{\top} {\bf \widetilde{A}} + \lambda \widetilde{\bf W}_e \widetilde{\bf W}_e) {\bm {\mathcal{X}}_g} -{\bf \widetilde{A}}^{\top} {\bf d}_g.
	\end{align}
\end{linenomath*}

\subsection{Alternating Coupled Inversion Workflow}
\label{subsec:Workflow}
\begin{figure}
	\noindent\includegraphics[width=\textwidth]{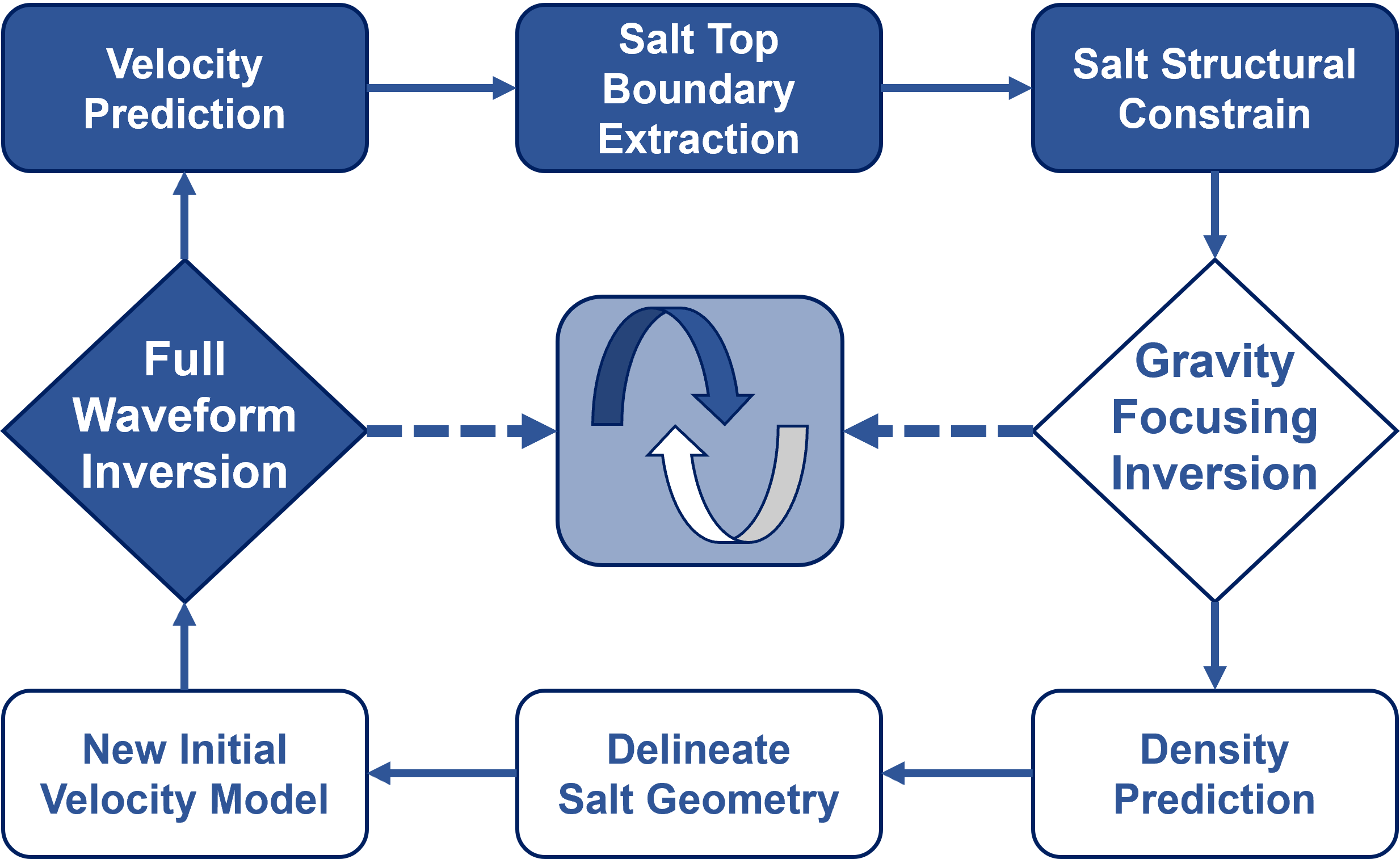}
	\caption{Flowchart of alternating coupled inversion.}
	\label{fg:flowchart}
\end{figure}

Based on the acoustic wave equation \cite{alford1974accuracy} and the focusing inversion algorithm \cite{portniaguine1999focusing}, we conduct 2-D multi-scale constant density FWI for P-wave velocity and GFI for density in salt dome, respectively. By employing the proposed alternating coupled inversion method (see Figure \ref{fg:flowchart}), FWI and GFI are alternately constrained, jointly improving the inversion quality of both P-wave velocity and density. This method is more robust and computationally less expensive than attempting to invert for velocity and density simultaneously, and it allows for better individual control over each problem. This workflow consists of two stages (Stage-I and Stage-II), with specific details as follows:

1) First, a horizontally layered velocity model is used as the starting point for FWI-Stage-I. The resulting velocity inversion is generally poor but can effectively recover the salt top boundary.  

2) The salt top boundary is extracted by calculating the maximum vertical velocity change for each trace, and this structural information is incorporated as geological constraint into GFI-Stage-I.  

3) The density updates are restricted to the region below the salt top boundary in GFI. Leveraging the inherent skin effect of gravity data, a well compacted and sharply focused density model can be obtained, which consistent with its top boundary.  

4) The salt dome shape delineated from GFI-Stage-I result is assigned a known salt velocity value, producing a large-scale but coarse salt velocity model. This model is then embedded into the original horizontally layered velocity model to serve as the starting point for FWI-Stage-II.

5) The more accurate top boundary information provided by FWI-Stage-II is then used as a new structural constraint in GFI-Stage-II.

Overall, FWI provides a reliable structural constraint for GFI, while GFI supplies a more accurate initial salt model for FWI.

\section{Numerical Examples}
\label{sec:examples}
We apply the proposed method to modified BP salt model and SEG/EAGE salt model with 6\% Gaussian noise. Table \ref{table:frequency_bands} provides the frequency bands and iterations for the multi-scale FWI in both model tests.

\begin{table}
	\caption{Parameters of multi-scale FWI in two model tests.}\label{table:frequency_bands}
	\setlength{\tabcolsep}{10pt}
	\centering
	\begin{tabular}{l c c c c c c}\toprule
		\textbf{BP2004 model} & Stage1 & Stage2 & Stage3 & Stage4 & Stage5 & -\\
		\hline
		Frequency (Hz) & < 5 & < 7 & < 10 & < 15 & < 30 & -\\
		Iterations (FWI-Stage-I) & 100 & 100 & 100 & 100 & 100 & -\\
		Iterations (FWI-Stage-II) & 150 & 150 & 150 & 150 & 150 & -\\
		\toprule
		\textbf{SEG/EAGE model} & Stage1 & Stage2 & Stage3 & Stage4 & Stage5 & Stage6\\
		\hline
		Frequency (Hz) & < 5 & < 7 & < 9 & < 11 & < 15 & < 30\\
		Iterations (FWI-Stage-I) & 100 & 100 & 100 & 100 & 100 & 100\\
		Iterations (FWI-Stage-II) & 100 & 200 & 200 & 200 & 200 & 200\\
		\bottomrule
	\end{tabular}
\end{table}

\subsection{Modified BP Salt Model}
\label{subsec:BP}
The modified BP salt model is derived from the center part of the original BP model and features a deeply rooted salt body. The main challenge here lies in velocity inversion of the salt body with steep dips \cite{billette20052004}. The model consists of 500 grid points in the horizontal direction and 150 grid points in the vertical direction, with a spatial sampling interval of 20 m in both directions. 

\begin{figure}
	\noindent\includegraphics[width=\textwidth]{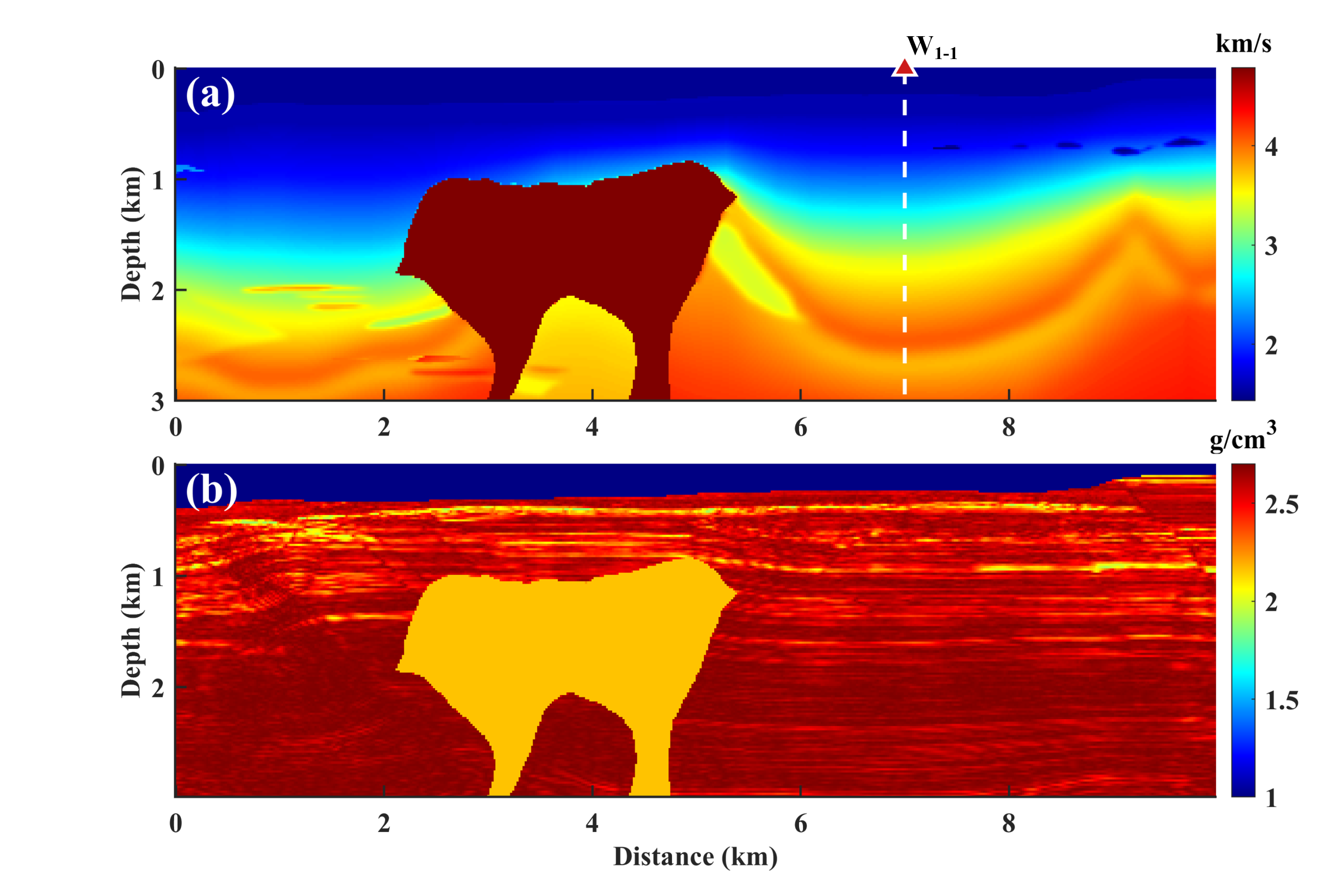}
	\caption{(a) True velocity, and (b) true density of the modified BP salt model. $\rm {W}_{1-1}$ is a hypothetical well used to construct the initial layered model for FWI-Stage-I.}
	\label{fg:BP_true}
\end{figure}

\begin{figure}
	\noindent\includegraphics[width=\textwidth]{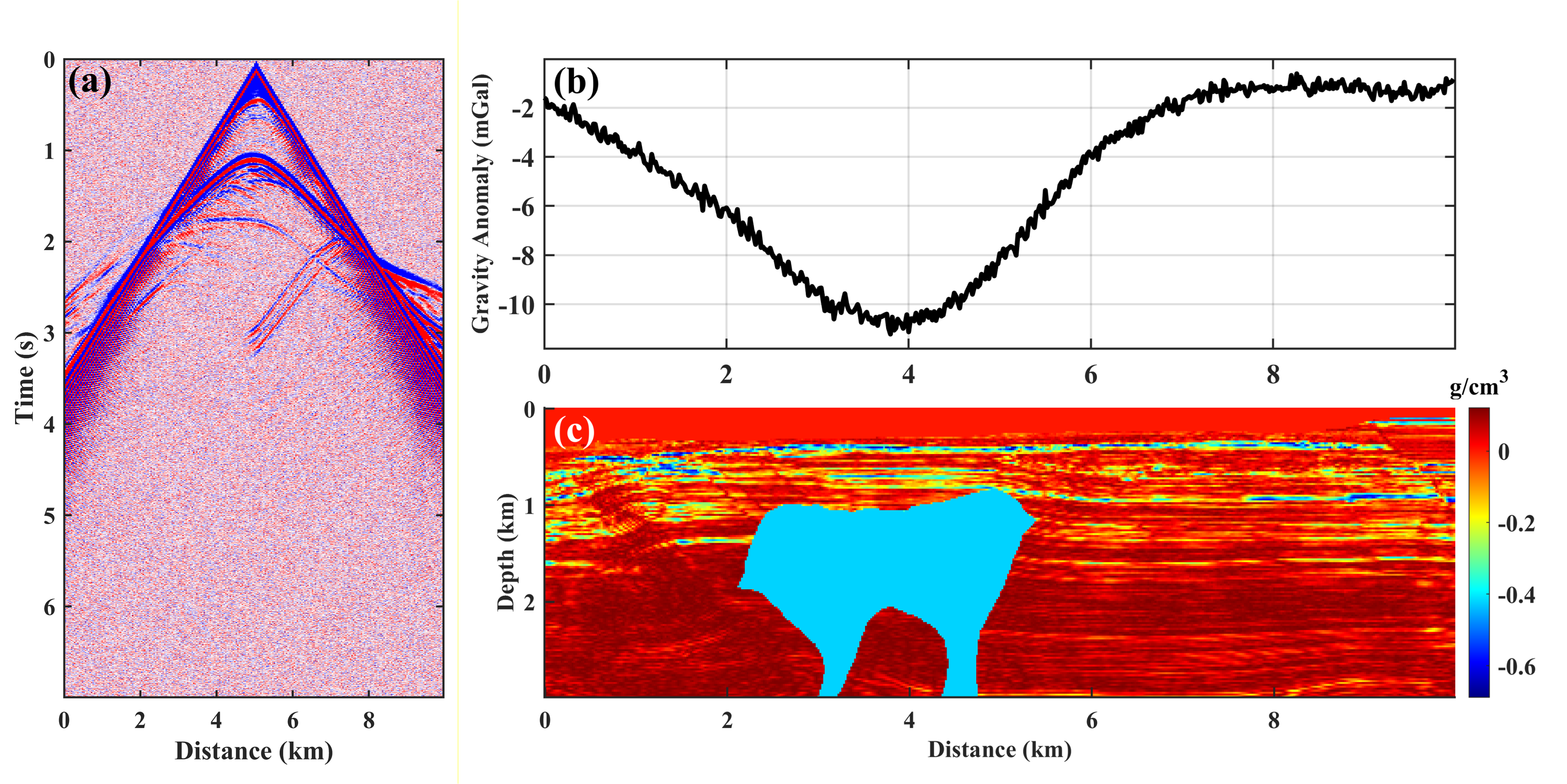}
	\caption{Measurement data of the modified BP salt model. The color scale of waveform data is clipped  between its 5\% and 95\% quantiles for enhanced visualization. (a) Waveform data for the 25 th source of 50 with 6\% Gaussian noise, (b) gravity data $\bm g_{\rm z}$ with 6\% Gaussian noise, (c) residual density $\bm \rho_{\rm salt}$ of the salt body.}
	\label{fg:BP_data}
\end{figure}

Figure \ref{fg:BP_true}a shows the true velocity model, with sources and receivers uniformly distributed along the surface. There is a total of 50 sources and 250 receivers. A Ricker wavelet with a peak frequency of 10 Hz is employed as the source function. The total recording time consists of 3,500 time samples with a sampling interval of 2 ms. Figure \ref{fg:BP_data}a shows the waveform data for the 25 th source of 50 with 6\% Gaussian noise. Figure \ref{fg:BP_true}b shows the true density model $\bm \rho_{\rm true}$. Since the background density distribution is relatively uniform, we treat it as a constant and compute the Bouguer gravity anomaly for the modified BP salt model. The specific procedure is as follows:

1) Perform forward modeling to compute the gravity response $\bm g_{\rm true}$ corresponding to $\bm \rho_{\rm true}$, which includes the full contribution of both the salt body and the surrounding sedimentary layers. 
 
2) Remove the salt body via masking and calculate the average background density $\bm \rho_{\rm bg} = \rm {2.58\ g/cm^3}$ using only the sedimentary layers. Assuming a homogeneous background, construct a uniform background model using $\bm \rho_{\rm bg}$ without salt body and perform forward modeling to obtain $\bm g_{\rm bg}$.

3) Obtain the Bouguer gravity anomaly $\bm g_{\rm z}$, which arises almost entirely from the residual density $\bm \rho_{\rm salt} = \rm {-0.44\ g/cm^3}$ of the salt body (see Figure \ref{fg:BP_data}c), by computing $\bm g_{\rm true} - \bm g_{\rm bg}$.

The gravity response caused by the water layer is removed in the above procedure. This approach of extracting the gravity response induced by a local anomalous body through background field stripping is widely applied in gravity inversion for salt dome \cite{lin2019gramian, chen20233, xian2024recovering}. In a strict sense, $\bm g_{\rm z}$ represents the approximate Bouguer gravity anomaly of the salt body. However, since the contribution to $\bm g_{\rm z}$ primarily comes from the residual density of the salt body and the background sediments are assumed to be homogeneous, the influence of local variations in the shallow background on $\bm g_{\rm z}$ can be approximately neglected.

\begin{figure}
	\noindent\includegraphics[width=\textwidth]{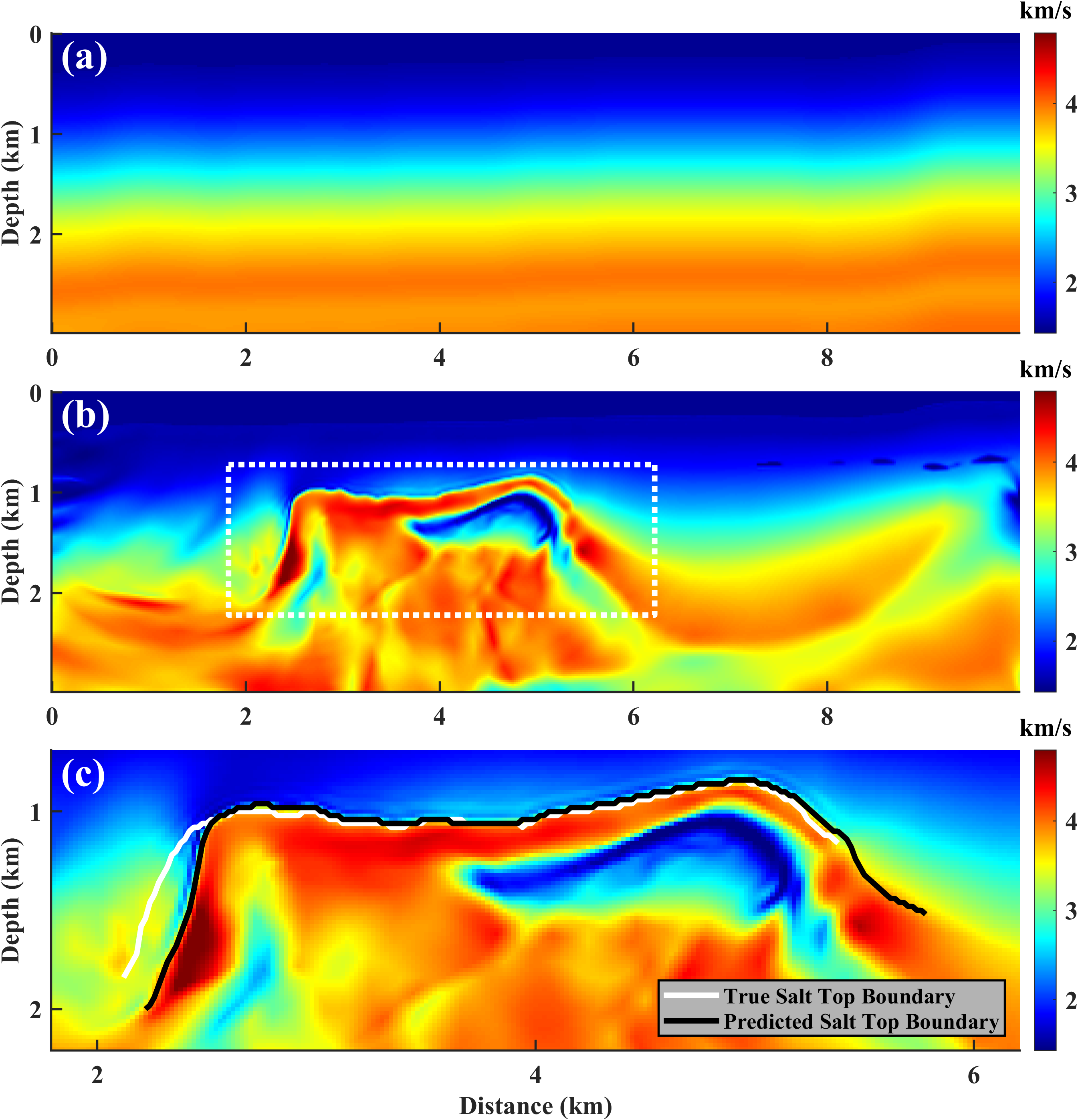}
	\caption{FWI-Stage-I of the modified BP salt model. (a) initial velocity model, (b) inversion result, and the white dashed line represents the locally enlarged range in (c) to demonstrate the boundary extraction effect.}
	\label{fg:BP_fwi_stage1}
\end{figure}

\begin{figure}
	\noindent\includegraphics[width=\textwidth]{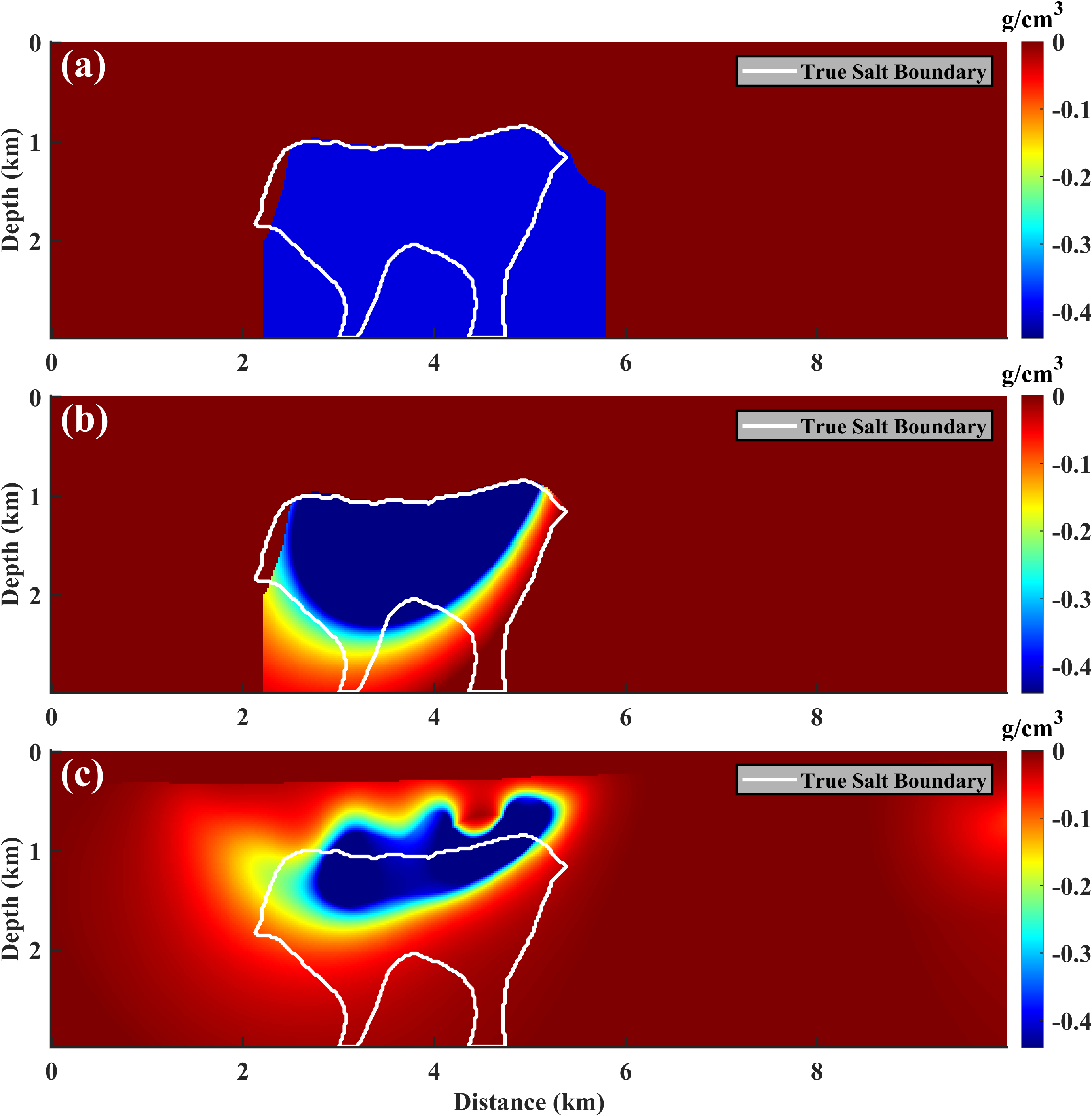}
	\caption{(a) Initial density model, and (b) inversion result of GFI-Stage-I. (c) GFI result without salt top structural constrain.}
	\label{fg:BP_gra_inv}
\end{figure}

We assume the existence of a known well log $\rm {W}_{1-1}$ within the survey area (indicated by the white dashed line in Figure \ref{fg:BP_true}a). The sedimentary layer velocities extracted from the well log data are successively extended along the seabed and smoothed using a Gaussian filter. This process generates the initial velocity model for FWI-Stage-I (see Figure \ref{fg:BP_fwi_stage1}a). Figure \ref{fg:BP_fwi_stage1}b presents the final inversion result from FWI-Stage-I. Due to the significant discrepancy between the initial model and the true model, the inversion result may falls into a local minimum. Although the salt body is not recovered well, the inversion result reasonably delineates its top boundary. When delineating this top boundary by computing the maximum vertical velocity change for each trace, we deliberately introduce some erroneous boundary segments to test the robustness of the algorithm. Ultimately, the black solid line in Figure \ref{fg:BP_fwi_stage1}c indicates the predicted salt top boundary, which we then use as a geological constraint to construct the initial density model (see Figure \ref{fg:BP_gra_inv}a) and to constrain GFI-Stage-I. Figure \ref{fg:BP_gra_inv}b presents the inversion result from the GFI-Stage-I. Compared to the GFI result without structural constraints (see Figure \ref{fg:BP_gra_inv}c), the incorporation of salt top structural constraints significantly compresses the solution space, reduces non-uniqueness, and mitigates the inherent low depth resolution problem in gravity inversion.

\begin{figure}
	\noindent\includegraphics[width=\textwidth]{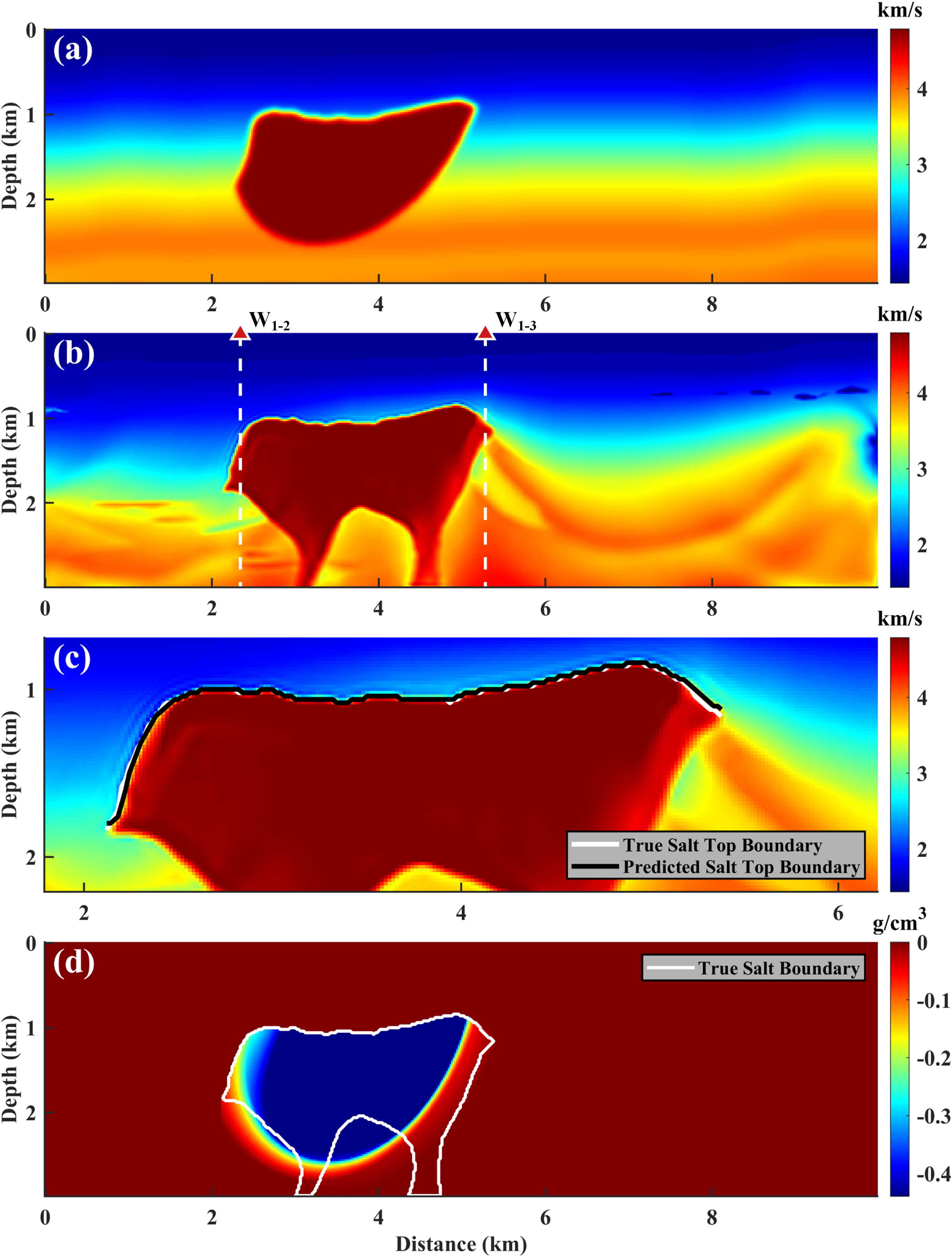}
	\caption{(a) Initial velocity model, (b) inversion result, and (c) the same locally enlarged range as in Figure \ref{fg:BP_fwi_stage1}b is used to demonstrate the boundary extraction effect of FWI-Stage-II. (d) Density inversion result of GFI-Stage-II. $\rm {W}_{1-2}$ and $\rm {W}_{1-3}$ are hypothetical wells assumed for the analysis of FWI-Stage-II.}
	\label{fg:BP_stage2}
\end{figure}

\begin{figure}
	\noindent\includegraphics[width=\textwidth]{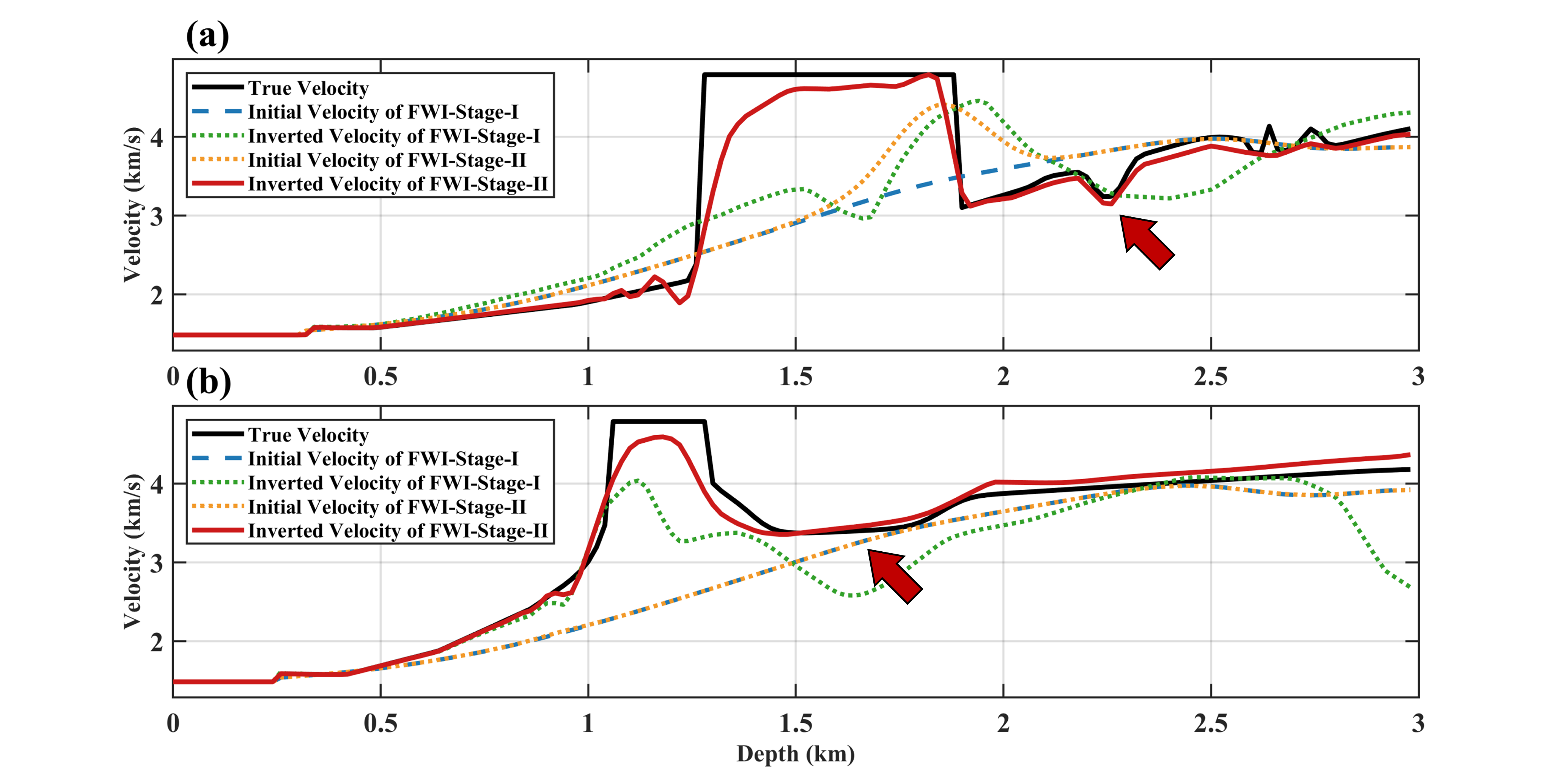}
	\caption{Comparison of velocity for the modified BP salt model at (a) $\rm {W}_{1-2}$ (2.34 km), and (b) $\rm {W}_{1-3}$ (5.28 km).}
	\label{fg:BP_curves}
\end{figure}

In delineating the salt body geometry, we identify region in the density inversion result shown in Figure \ref{fg:BP_gra_inv}b with values less than $\rm {-0.2\ g/cm^3}$ as the salt body. The known salt body velocity is assigned to these regions, which are then embedded into the horizontally layered velocity model from FWI-Stage-I. The resulting model serves as the initial velocity model for FWI-Stage-II (see Figure \ref{fg:BP_stage2}a). Figure \ref{fg:BP_stage2}b shows the final estimated velocity. The GFI provides a better initial model for FWI, enabling successful reconstruction of the salt body and background structures in well-illuminated regions. Notably, at the locations of wells $\rm {W}_{1-2}$ and $\rm {W}_{1-3}$, the low-velocity zones along the salt flanks correspond to potential hydrocarbon traps (indicated by red arrows in Figure \ref{fg:BP_curves}), FWI-Stage-II successfully recovers their characteristics. Figure \ref{fg:BP_stage2}c presents a more refined delineation of the salt top obtained from FWI-Stage-II, which is then used as a new structural constraint for GFI-Stage-II, leading to the more compact and reasonable salt density inversion result shown in Figure \ref{fg:BP_stage2}d.

\subsection{Modified SEG/EAGE Salt Model}
\label{subsec:SEG}
\begin{figure}
	\noindent\includegraphics[width=\textwidth]{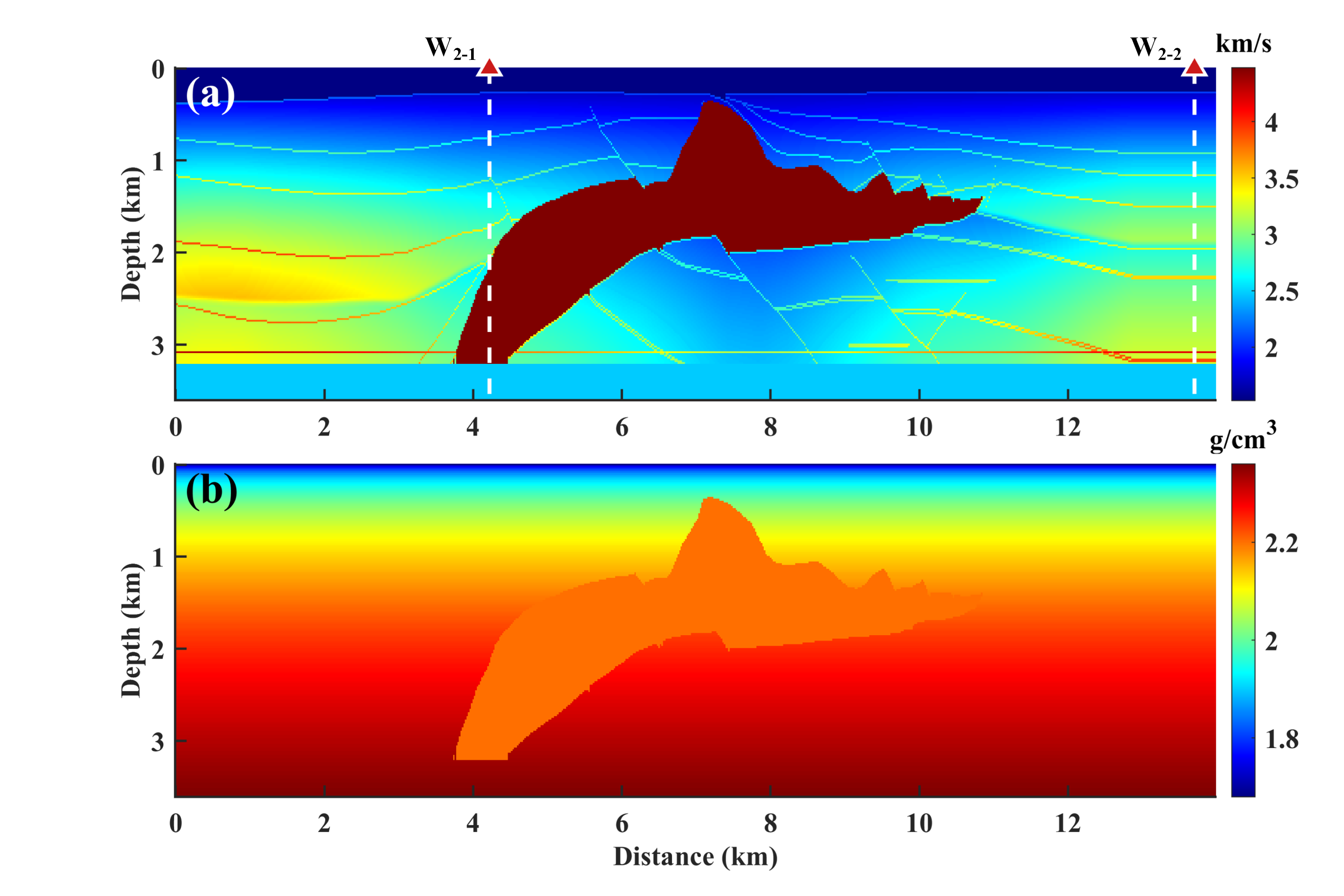}
	\caption{(a) True velocity, and (b) true density of the modified SEG/EAGE salt model. $\rm {W}_{2-1}$ and $\rm {W}_{2-2}$ are hypothetical wells assumed for constructing the initial horizontally layered models in FWI-Stage-I (W2-1) and FWI-Stage-II (W2-2), respectively.}
	\label{fg:SEG_true}
\end{figure}

The modified SEG/EAGE salt model (see Figure \ref{fg:SEG_true}a) is derived from one of the most classic slices of the 3-D SEG/EAGE salt model. It is observed that a massive piercing-type salt diapir is located in the central part of this slice. Due to the strong velocity variations in both horizontal and vertical directions, as well as the shielding effect caused by the large salt body, velocity inversion of the salt body and subsalt structures in this region poses a significant challenge. The model consists of 700 grid points in the horizontal direction and 181 grid points in the vertical direction, with a spatial sampling interval of 20 m in both directions. 

\begin{figure}
	\noindent\includegraphics[width=\textwidth]{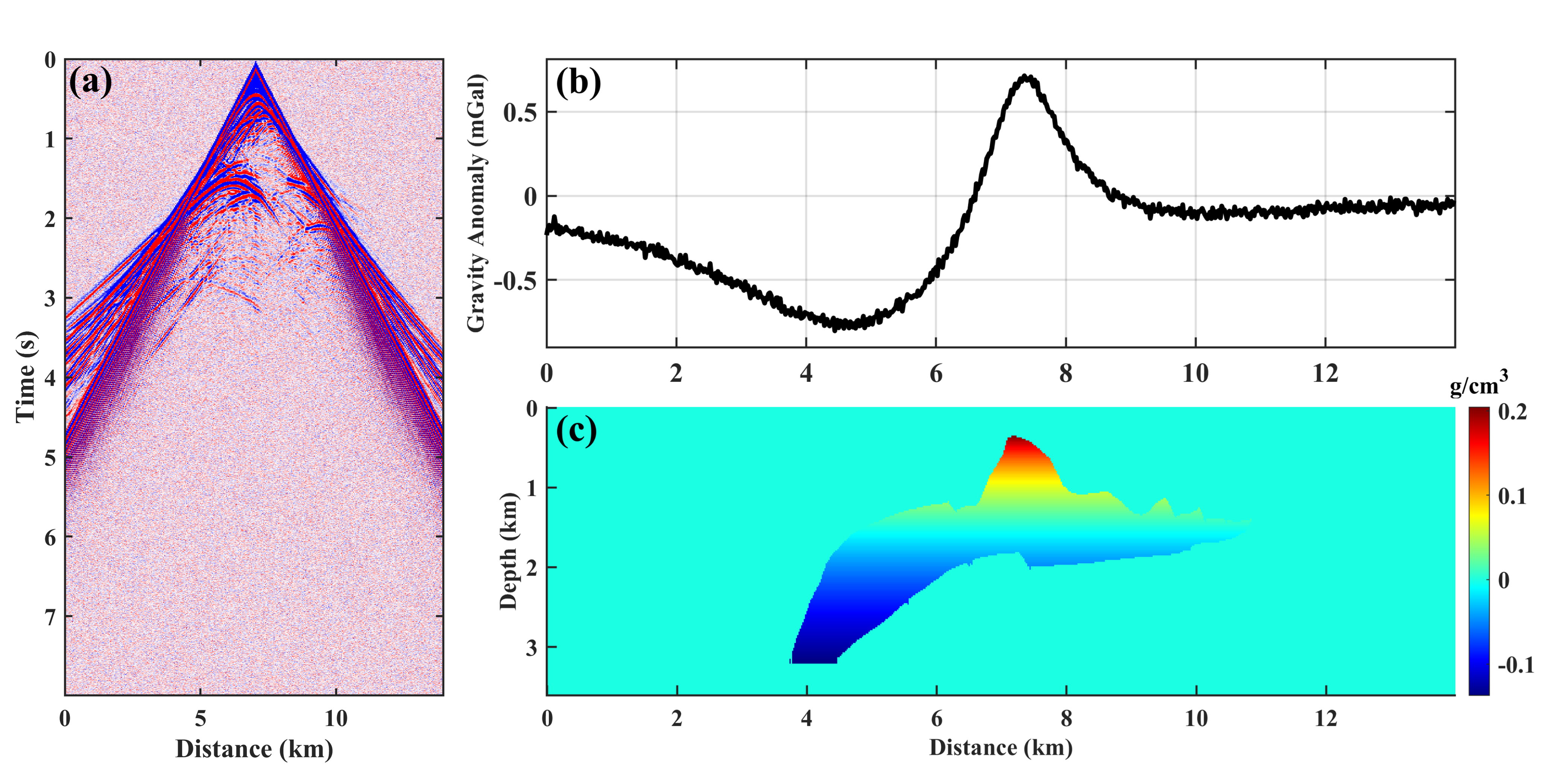}
	\caption{Measurement data of the modified SEG/EAGE salt model. The color scale of waveform data is clipped  between its 5\% and 95\% quantiles for enhanced visualization. (a) Waveform data for the 35 th source of 70 with 6\% Gaussian noise, (b) gravity data $\bm g_{\rm z}$ with 6\% Gaussian noise, (c) residual density $\bm \rho_{\rm salt}$ of the salt body.}
	\label{fg:SEG_data}
\end{figure}

For the acquisition setup, sources and receivers are uniformly distributed along the model surface, with a total of 70 sources and 350 receivers. A Ricker wavelet with a peak frequency of 10 Hz is employed as the source function. The total recording time consists of 4,000 time samples with a sampling interval of 2 ms. Figure \ref{fg:SEG_data}a shows the waveform data for the 35 th source of 70 with 6\% Gaussian noise. Unlike the modified BP salt model, where the density background is assumed constant, the density model for the modified SEG/EAGE salt model takes into account the compaction trend of the background density. This is manifested as a density contrast that varies with depth. This poses a significant challenge for GFI in salt dome, particularly in the null zone (at approximately 1.5 km depth in this study model) where the salt density matches that of the background sediments, resulting in a lack of gravity signal response \cite{krahenbuhl2006inversion,li2016level,wei2023quantifying}. The salt density is set to $\rm {2.2\ g/cm^3}$, while the background sedimentary density varies with depth according to relation $\rho = 1.4 + 0.172\rm {z}^{0.21}$ \cite{hudec2009paradox}, where $\rm z$ indicates the depth. The true density model is shown in Figure \ref{fg:SEG_true}b. The calculation steps for the Bouguer gravity anomaly of this salt body are similar to those used in the modified BP salt model. The resulting Bouguer anomaly with 6\% Gaussian noise and residual density of the salt body are shown in Figures \ref{fg:SEG_data}b and \ref{fg:SEG_data}c, respectively.

\begin{figure}
	\noindent\includegraphics[width=\textwidth]{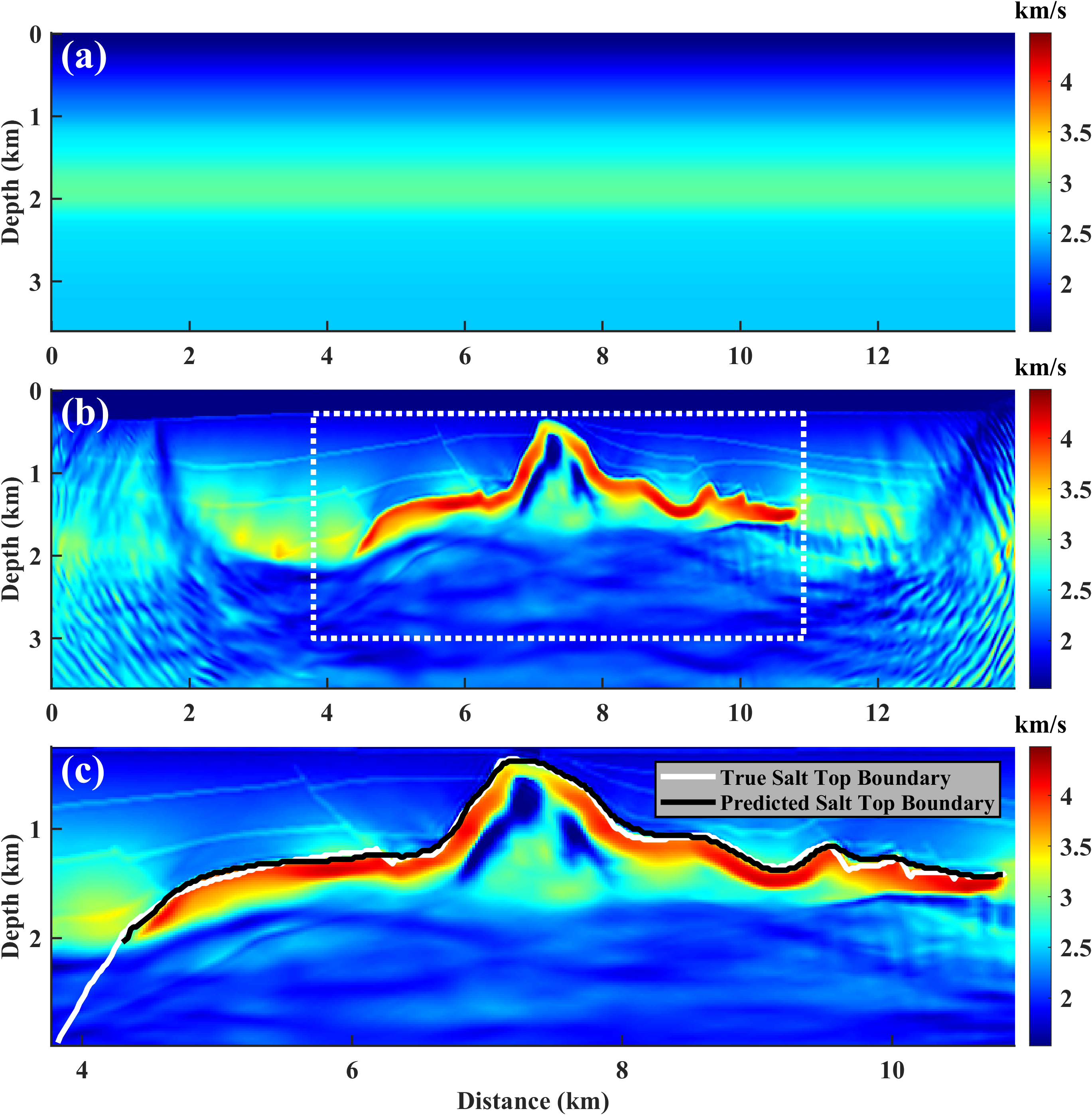}
	\caption{FWI-Stage-I (W2-1) of the modified SEG/EAGE salt model. (a) initial velocity model, (b) inversion result, and the white dashed line represents the locally enlarged range in (c) to demonstrate the boundary extraction effect.}
	\label{fg:SEG_fwi_stage1}
\end{figure}

\begin{figure}
	\noindent\includegraphics[width=\textwidth]{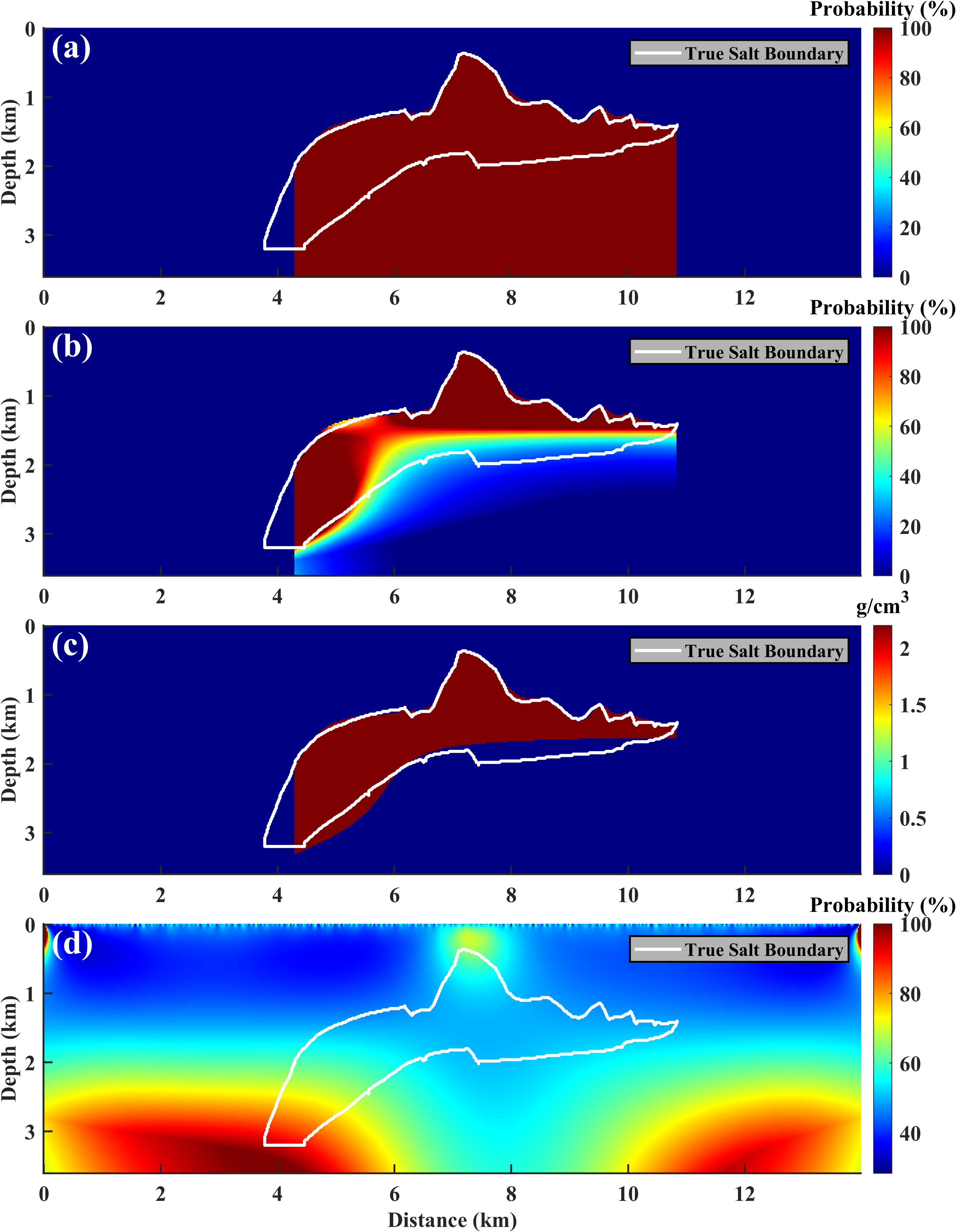}
	\caption{(a) Initial probability model, (b) predicted probability result, and (c) the salt body delineated by regions with a predicted probability greater than 40\% of GFI-Stage-I (W2-1). (d) GFI result without salt top structural constrain.}
	\label{fg:SEG_gra_inv}
\end{figure}

We extract stratigraphic velocity information from the hypothetical well $\rm {W}_{2-1}$, remove the velocity of the salt body, interpolate the values, and successively extend them along the seabed to obtain the initial velocity model after Gaussian smoothing (see Figure \ref{fg:SEG_fwi_stage1}a), which serves as the starting point for FWI-Stage-I (W2-1). Although the inversion result exhibits typical cycle-skipping features due to the poor quality of the initial model, such as nonphysical low-velocity zones below the top boundary of the salt body, the top boundary itself is still relatively well defined. As shown in Figure \ref{fg:SEG_gra_inv}a, in constructing the initial probability model for GFI-Stage-I (W2-1), we assume that all regions below the predicted salt top boundary (black solid line in Figure \ref{fg:SEG_fwi_stage1}c) belong to the salt body. Based on this assumption, GFI accounting for the depth varying density contrast is performed, and the resulting probability prediction is presented in Figure \ref{fg:SEG_gra_inv}b. Regions with a predicted probability greater than 40\% are considered as the interpreted salt body. Figure \ref{fg:SEG_gra_inv}d shows the GFI result without salt top structural constraint. Due to the presence of the null zone, the predicted probability of salt body increases with distance from the null zone, causing the inversion result to split into two disconnected parts with no effective information. In contrast, with the introduction of salt top boundary provided by FWI-Stage-I (W2-1), the result becomes focused and compact. Although the influence of the null zone remains, it is significantly reduced.

\begin{figure}
	\noindent\includegraphics[width=0.85\textwidth]{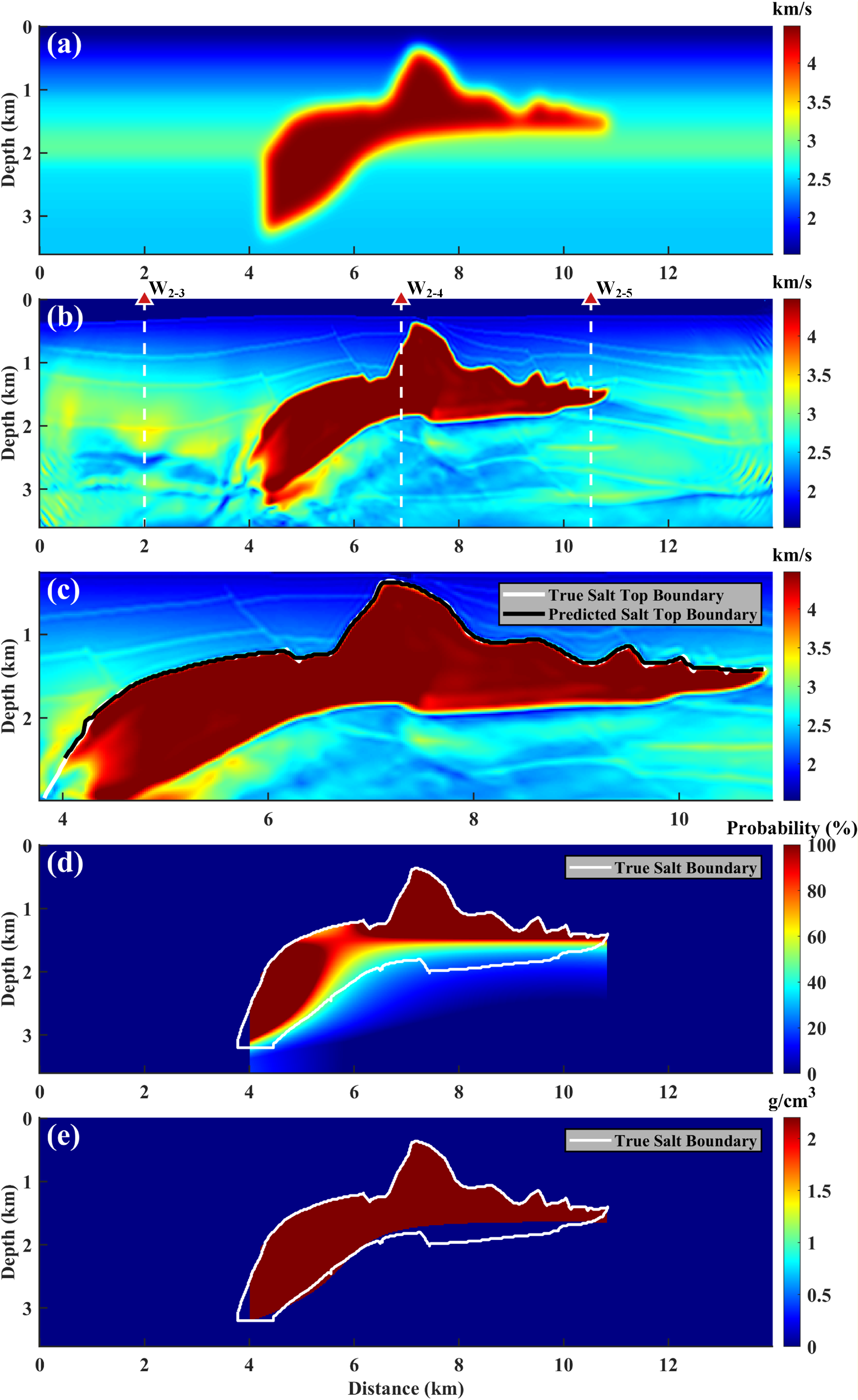}
	\caption{(a) Initial velocity model, (b) inversion result, and (c) the same locally enlarged range as in Figure \ref{fg:SEG_fwi_stage1}b is used to demonstrate the boundary extraction effect of FWI-Stage-II (W2-1). (d) Predicted probability result, and (e) the salt body delineated by regions with a predicted probability greater than 40\% of GFI-Stage-II (W2-1). $\rm {W}_{2-3}$, $\rm {W}_{2-4}$, and $\rm {W}_{2-5}$ are hypothetical wells assumed for the analysis of FWI-Stage-II (W2-1) and FWI-Stage-II (W2-2).}
	\label{fg:SEG_stage2}
\end{figure}

The salt body delineated in Figure \ref{fg:SEG_gra_inv}c is assigned the known salt dome velocity and then embedded into the horizontally layered velocity model from FWI-Stage-I (W2-1) to construct the initial model for FWI-Stage-II (W2-1) (see Figure \ref{fg:SEG_stage2}a). The final estimated velocity is shown in Figure \ref{fg:SEG_stage2}b. Benefiting from the salt body delineation provided by GFI-Stage-I (W2-1), FWI-Stage-II (W2-1) achieves a better reconstruction of both the salt body and subsalt velocities. Its more accurate delineation of the salt top further enhances the salt body positioning prediction and density inversion in GFI-Stage-II (W2-1) (see Figures \ref{fg:SEG_stage2}d and \ref{fg:SEG_stage2}e).

\begin{figure}
	\noindent\includegraphics[width=\textwidth]{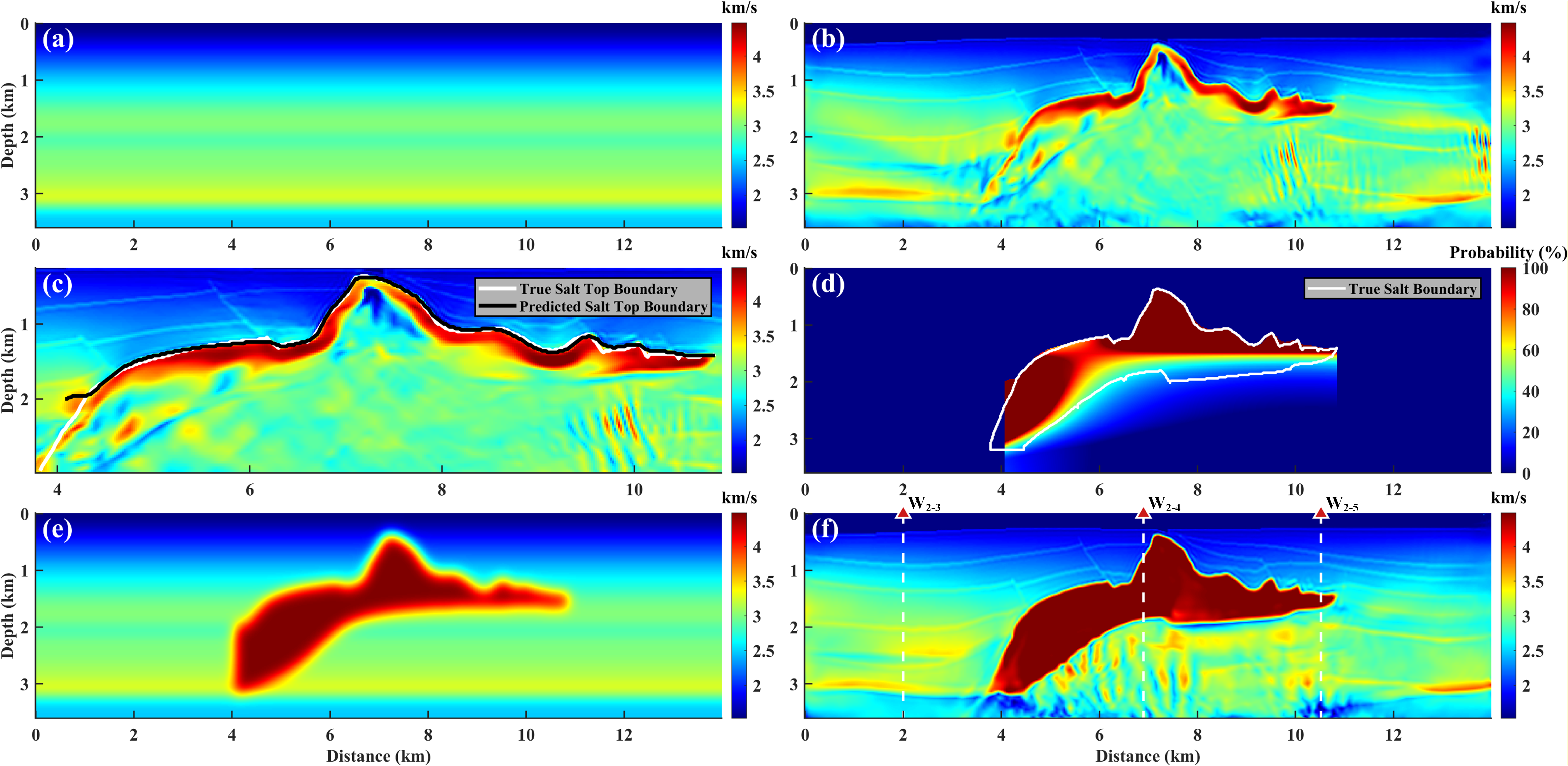}
	\caption{(a) Initial velocity model, (b) inversion result, and (c) the same locally enlarged range as in Figure \ref{fg:SEG_fwi_stage1}b is used to demonstrate the boundary extraction effect of FWI-Stage-I (W2-2). (d) Predicted probability result of GFI-Stage-I (W2-2). (e) Initial velocity model and (f) inversion result of FWI-Stage-II (W2-2).}
	\label{fg:SEG_s1}
\end{figure}

\begin{figure}
	\noindent\includegraphics[width=\textwidth]{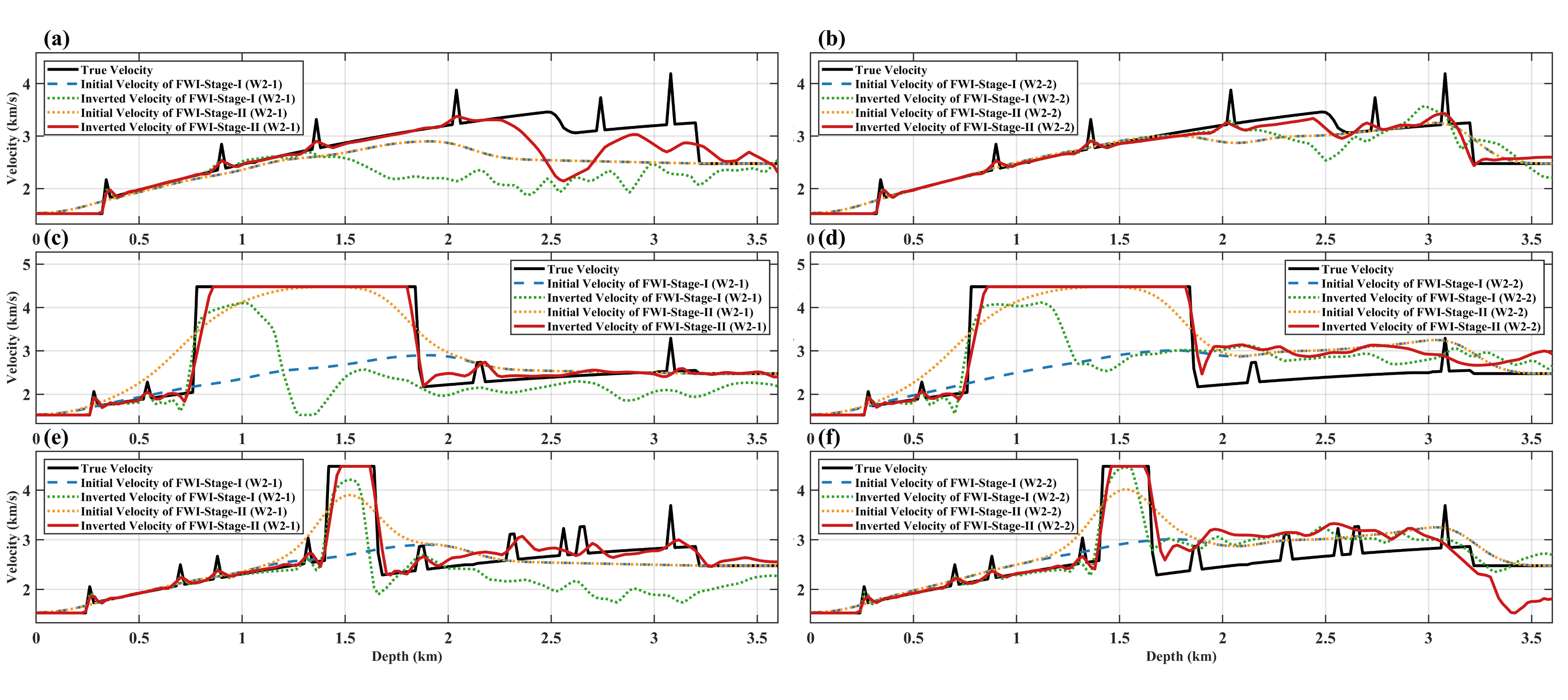}
	\caption{Comparison of velocity profiles from FWI-Stage-I/II (W2-1) and FWI-Stage-I/II (W2-2) at well (a) and (b) $\rm {W}_{2-3}$ (2.00 km), (c) and (d) $\rm {W}_{2-4}$ (6.90 km), and (e) and (f) $\rm {W}_{2-5}$ (10.52 km) in the modified SEG/EAGE salt model.}
	\label{fg:SEG_curves}
\end{figure}

However, it is worth discussing the nonphysical low-velocity zone observed at a depth of approximately 2.5 km near well $\rm {W}_{2-3}$ in Figure \ref{fg:SEG_stage2}b. The primary reason for this artifact is attributed to the initial model having low velocities in this region that deviate significantly from the true model, combined with poor illumination, which makes the inversion more prone to cycle skipping and falling into local minima. Using the stratigraphic velocity from well $\rm {W}_{2-1}$ to construct the background model leads to relatively low velocities in the deeper section. For comparison, the stratigraphic velocity from well $\rm {W}_{2-2}$ is adopted to build an alternative background model with higher deep velocities (Figure \ref{fg:SEG_s1}a). Figure \ref{fg:SEG_s1} presents the detailed results of FWI-Stage-I/II (W2-2), where the delineation of salt top and its constraining effect on GFI results are largely consistent with those of FWI-Stage-I (W2-1). The most notable difference is observed in Figure \ref{fg:SEG_s1}f. Although the higher background velocity in the initial model leads to a more accurate velocity inversion result near well $\rm {W}_{2-3}$, it cannot be overlooked that the recovery of subsalt velocities deteriorates (see Figure \ref{fg:SEG_curves}). This outcome is to be expected. The shielding effect of the massive salt body significantly reduces the number of effective wave paths that penetrate the salt and illuminate the subsalt region. This results in lower sensitivity of the data to subsalt velocity perturbations. In FWI-Stage-I (W2-2), the initial velocity model in the subsalt region deviates considerably from the true model, which increases waveform phase mismatch and further exacerbates the non-uniqueness of FWI.

The comparison between FWI-Stage-I/II (W2-1) and FWI-Stage-I/II (W2-2) further demonstrates that constant density acoustic FWI is highly sensitive to the initial background velocity model. In industrial-scale FWI, to mitigate its ill-posedness, velocity model building (VMB) is typically a highly refined and strongly interactive process that requires extensive geological prior knowledge and human expertise to progressively establish a suitable initial model, providing a more reliable starting point for subsequent FWI \cite{jones2010introduction, leveille2011subsalt, wang2019full, jiao2008practical}. In contrast, this study focuses on validating the feasibility of an alternating iterative framework between GFI and FWI in complex salt environments. As a result, a more simplified prior input (i.e., single-well velocity information and gravity constraints) is adopted in the VMB process, which carries the risk of compromising the accuracy of the background velocity in localized regions and introducing certain artifacts. Nevertheless, the experimental results indicate that even this simplified workflow can achieve interpretable improvements in overall trends and key structural delineations, such as salt body identification, salt top boundary definition, even subsalt velocity reconstruction. This demonstrates a viable pathway for multi-physics data inversion under limited prior information.

\section{Conclusion}
In this study, we propose a multi-physics alternating coupled inversion method based on FWI and GFI. The method exploits the complementary advantages of full waveform and gravity data, which provide information at different scales. Through an alternating iterative procedure, it progressively improves the reliability of the inverted velocity and density models. The main conclusions are as follows:

1) FWI and GFI exhibit strong complementarity. FWI provides high-resolution velocity structures and reliable salt top structural information, while GFI stably recovers a compact density distribution of salt body and supplies long-wavelength structural constraints for seismic exploration.

2) The proposed alternating coupled inversion strategy constructs a macroscopic salt model from GFI result, significantly improving the quality of initial model for FWI, thereby mitigating cycle-skipping issues and enhancing inversion stability.

3) Incorporating a depth-varying density contrast into GFI better reflects realistic geological conditions. Under the constraint of salt top structure from FWI result, this approach effectively alleviates the null-zone and annihilation problems, leading to more geologically plausible density inversion results.

Numerical experiments conducted on modified BP salt model and SEG/EAGE salt model demonstrate that the proposed method yields accurate salt geometries and velocity structures compared to the conventional FWI and GFI method. This validates its effectiveness for velocity and density inversion in complex salt dome. Overall, the alternating coupled inversion method presented in this work offers a new paradigm for multi-physics inversion in salt dome.

%\section{Appendix A}
%\label{sec:appendix_a}

%\begin{table}
%	\caption{Parameters of multi-scale FWI in two model tests.}\label{table:GFI_parameters}
%	\setlength{\tabcolsep}{10pt}
%	\centering
%	\begin{tabular}{l c c c}\toprule
	%		\textbf{BP2004 model} & e & $\lambda$ & Iterations\\
	%		\hline
	%		GFI-Stage-I & 0.1 & < 7 & 100\\
	%		GFI-Stage-II & 0.1 & 100 & 100\\
	%		Iterations (Stage-II) & 150 & 150 &\\
	%		\toprule
	%		\textbf{SEG/EAGE model} & Stage1 & Stage2 & Stage3\\
	%		\hline
	%		Frequency (Hz) & < 5 & < 7 & < 9\\
	%		Iterations (Stage-I) & 100 & 100 & 100\\
	%		Iterations (Stage-II) & 100 & 200 & 200\\
	%		\bottomrule
	%	\end{tabular}
%\end{table}

\bibliographystyle{unsrt}  
\bibliography{references}

\end{document}